# Seminar Innovation Management

**Pattern Recognition Lab,
Friedrich-Alexander-Universität Erlangen-Nürnberg**

**and**

**Sony Intellectual Property Europe**

Winter term 2016/17

# Contents

# 1 AR/VR Glasses with Medical Applications

Authors: Gerd Häusler, Aleksandra Milczarek, Markus Schreiter, Thomas Kästner, Florian Willomitzer, Andreas Maier

## 1.1 Problem Statement

For the prevention of eye diseases an early diagnosis and treatment is necessary. This requires simple means for detection. Multiple eye malfunction / diseases as well as personal parameters (e.g. fatigue), which can be detected by eye movements, are not frequently monitored. For some diseases, which cannot be treated (e.g. color blindness), augmented reality approaches can be used for the compensation or even to support the users without diseases in critical vision situations (e.g. low light, fog). No current method or device that combines both augmented reality (AR) and virtually reality (VR). No device for the gradual reduction of the transparency of glasses for each eye separately. No current method or device that stimulates the eye based on the analysis of cardiac cycle signals.

Disadvantages of the state of the art:

- Smart glasses do not support eye observation for medical diagnostics as well as long term observations of the human eye.

- Projection of augmented reality by smart glasses does not exist for both eyes simultaneously.

- No local zoom function that considers eye and head movement.

- Transparency of the glasses cannot be gradually reduced. This would enable sunglasses as well as switching between virtually reality (VR) and augmented reality (AR) for instance for zoom function.

- No real time color correction/ enhancement as well as compensation of color vision deficiency

- No vision enhancement based on AR due to cardiac cycle signals

## 1.2 Proposed Method

- Providing a method and wearable device for the detection and compensation of eye dysfunction and diseases. Eye motion is detected using eye tracking techniques.

- Visual feedback in daily life based on cardiac cycle signal (also for ordinary /healthy people). I.e. the system can inform users about stress and arousal and ask them to rest or cool down. Identification of excessive and sudden exposure to light will cause stress with can be identified and circumvented.

- System is able to work as AR and VR system at the same time. AR contrast can be improved using only partial / pixel-wise opacity. Pixels that come from the AR / VR world are no longer mixed with the real world signal. This results in a more realistic experience.

- System also analyzes the light sources with additional sensors and is able to render AR objects with realistic shading.

Setup:

- wearable device

- eye glass/es with adjustable transparency (e.g. e-paper / e-ink)

- projector/s for augmented reality (to the eye)

- camera/s to observe the eyes

- camera/s to observe the real world

- Sensor/s for cardiac cycle monitoring

Both eye movements are observed by separate cameras. Algorithmic evaluation of the eye movement for pre-diagnostics (e.g saccades, average motion, average speed, distribution of movements and speed, maximal motion) of possible diseases (e.g. Glaucoma, dementia, etc.) and other conditions of the user (fatigue, stress, fear, etc.). Material with adjustable transparency (e.g. FLCOS display) is embedded in each glass to enable dimming. Via augmented reality AND virtual reality approaches, people with eye malfunctions are supported in daily life. For VR, the glasses have to be completely dimmed. Some additional sensors (e.g. heartbeat-sensors) are embedded on the inside of the bails to support monitoring functions and can be used for adaption purposes.

Examples for possible usages:

- Edge enhancement in real time (e.g.: under low light condition)

- Amplification in low light situations

- Dimming in bright situations

- Magnification of the current field of view. Also with view-in-view mode or only magnification in partial areas of the field of view.

- Enabling VR by the gradually reduction of transparency to zero (for one or both glasses). Therefore virtual images are projected either in one or both eyes (e.g.: magnification of the current field of view -> zoom function, can be used for entertainment purpose, etc.)

- cardiac cycle signals, information from eye observation and mimic are measured and analyzed (e.g.: for detection of an acute stroke event, etc.). Thus for instance conclusions regarding diseases, the emotional state and mood can be made.

- Based on the interpretation of the measured signals the eye is stimulated (AR). For instance: dimming light in bright situations, Edge enhancement and Amplification in low light condition (in real time)

- increasing color intensity, changing color values adding object to the reality (AR applications e.g.: entertaining purpose, adding a 3D objects into the field of view [for instance: a Porsche in front of your garage, education purpose: overlaying reality with virtual images such as showing a dinosaur in the classroom], simulation of color blindness)

- Further usage of AR applications for increase the intensity of colors (entertaining application for fun)

- Color blindness compensation by overlaying intrinsic color with information that are visible to the subject (AR for medical usage). Kind and degree of color blindness is estimated from a dedicated test protocol that can be run in VR mode. AR mode compensates for color blindness by interpolating false colors into the scene.

- Method can also be extended to hyper spectral imaging where colors that cannot be differentiated by the eye (metamers) are in addition acquired by a hyper spectral camera mounted on the glasses. Metamers can then be displayed in false colors in the AR system.

## 1.3 Advantages of the Idea

- Support of eye observation for medical diagnostics as well as long term observations of the human eye. (on the principle of smart glasses the data are stored on an external device, also analyses can be outsourced)

- Projection of augmented reality by smart glasses simultaneously for both eyes. (stereo vision)

- Local zoom function that considers eye and head movement (by using eye tracking the focus point of the subject is detected, simultaneously focused object is enlarged and projected into the eye -> VR)

- Transparency of the glasses can be gradually reduced. This enables sunglasses as well as switching between virtually reality (VR) and augmented reality (AR) for instance for zoom function.

- Real time color correction or enhancement as well as compensation of color vision deficiency. Color blindness has to be determined in a VR calibration phase. (color blindness reduced for instance by overlaying the real world with virtual images that contain colors with changed color values; color enhancement is based on the same principle: overlaying reality with a virtual image projected in the eye)

- Recording of cardiac cycle signals and analyzation (usage for: e.g. vision enhancement based on AR, several diseases, etc.); (possible scenario: increased heart beat frequency is detected, furthermore mimic is analyzed, etc. -> conclusions on the emotional state -> timing the light/changing the color in order to enable quick relaxation to normal stress state)

- Combination with additional hardware components possible (e.g. IR camera - for night vision, thermography, etc.)

## 1.4 Detectability of the Idea in a Product

- Combination of AR and VR in one device

- Color correction/visual improvement based on AR

- Visual enhancement in daily life (zoom function, etc.)

# 2 Augmented Reality Eye Surgery

Author: Florian Schiffers

## 2.1 Problem Statement

Endoscopic surgery inside the eye is a difficult task due to two reasons. 1) Vision/health of the patient is constantly endanderegd during operation, and 2) inside-the-eye-image quality is bad since image retrieval is difficult. Estimation of position and orientation of the operating/imaging devices inside the eye is difficult for the surgeon. Thus, navigation inside the eye is tricky. Different imaging (2D and 3D) techniques of the eye are already state-of-the-art (OCT, fundus, CT, MRI). Currently no techniques using prior information about the eye is used in a setup for eye-surgeon guidance. Augmented reality methods for minimal invasive surgery is state-of-the-art, but no methods are designed specifically for inside eye-surgery.

## 2.2 Proposed Method

A novel setup for augmented reality for endoscopic eye surgery is presented. Additional patient information (obtained by prior 2D/3D imaging of the patient eye) is used to provide better real-time guidance for surgeons eventually leading to higher success rates in operations and enable new kind of eye operations, which are currently too risky for the patient. A 3D model of the patient eye is created in the operation planning phase. 3D modeling can be done by applying any technique leading to a 3D model of the human eye. Possible techniques are OCT, CT, MRI or topography measurement of the eye ball from the outside. Assuming a 3D model of the eye is given (see above), a virtual camera is now placed on the tip of the operating device (e.g. tweezers). If desired, the surgeon can choose any artificial position and orientation of the virtual camera. he 3D model is augmented with prior acquired images (e.g. fundus imaging) or in-vivo images during operation, for example from a second endoscope.

Further – on the fly – refinement of the 3D model is done by using real time information (e.g. by employing Structure-From-Motion techniques). Additional knowledge (like tumor segmentation, vessels) is augmented on the 3D model to provide further guidance. The additional information, which should be augmented, is displayed on an external device. For example: monitor or virtual reality glasses.

External or internal 3D tracking of the endoscopic devices and is mandatory. An external real time tracking of the location and pose of both, the eye and the operating devices inside the eye, are crucial. Any tracking providing information about position and orientation above can be used, among those are optical and robot-guided tracking for the endoscopic devices and optical (interferometry, triangulation, deflectromtry) techniques for the eye. Information of the position and pose of the operating device and eye is combined by registration techniques to have well defined coordinate system enabling software solution for augmented reality.

Additional information can conveyed to the surgeon if needed:

- Thickness of the different layers is segmented from the OCT and displayed as additional parameters

- Biological important structures (e.g. Vessels, Tumors) are highlighted during operation

- Pressure of the operating devices is measured by mounted device and displayed

- Distance to retina of the operating device is displayed as colored information. In case operating device is close to critical areas in the eye, feedback (optical or haptic) is given to warn the surgeon

Generated 3D models can printed out. Assuming a high quality artificial replica of the eye can be provided, the artificial eye can be used for training scenarios for (future) eye surgeons with our proposed setup.

Optionally:

- Endoscopic devices can be controlled by remote reducing the risk during operation since the surgeon is never free of errors or external distraction in the operating room

- The software solutions allows for easy integration into a fully automated setup using robots

## 2.3 Advantages of the Idea

The idea enables better visualization methods assisting the surgeon leads to higher operation success rates or even new operation methods. In addition, it enables better guidance of the endoscope and operating device.

## 2.4 Detectability of the Idea in a Product

Augmented reality approaches comprising a prior generated 3D Model are always displayed on an external device (e.g. monitor, virtual reality glasses). Real-time tracking device of endoscope and eye has to be present in the operation room

# 3 Game Diagnosis

Authors: Stefan Steidl, Temitope Paul Onanuga, Mathias Unberath, Florian Dötzer, Maike Stöve, Jonas Hajek

## 3.1 Problem Statement

While regular examinations of childrens eye health by a doctor are common today, the intervals in between them are of the order of a year or at least months. As changes in eye health can take place much faster than this, the need for early diagnosis arises. Furthermore, children often are not very cooperative during examinations at a doctor, potentially falsifying the results of subjective tests there.

  The idea shall offer constant monitoring of children's eye health at home. It allows parents to get notifications when their child's behavior changes in a way that could be connected to an impairment of their eyesight. It is not designed to replace the examinations at a doctor, but it shall offer the additional ability to react to health issues more quickly and have a visit at the doctor earlier then. Applications that test someones eyesight are widely available, but they usually rely on the active involvement of the child. Our solution provides passive testing, i.e. without the child even noticing it is being tested.

## 3.2 Proposed Method

The idea is to integrate hidden eyesight tests into video games. One can then create statistics about the child's success in different situations and how this changes over time. Similar ideas were proposed in US6808267B2, US8931905B2, EyeSpy 20/20.
As in US6808267B2, US8931905B2, US9433346B2, the necessary hardware includes:

- Computer (e.g. console, PC)

- Screen (of known size)

- Device that measures distance between a person and the screen

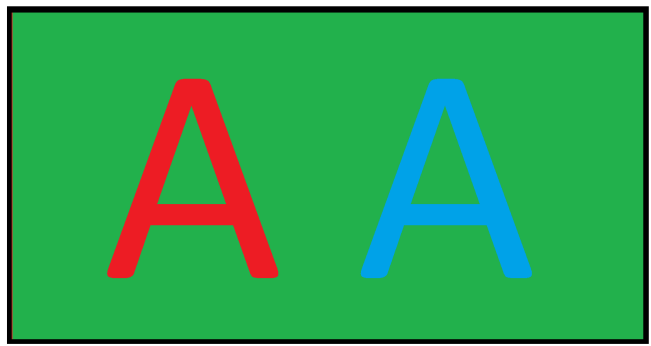

**Figure 1:** Features of different colours in front of a background

- Ambient light sensor

The computer and screen are standard technology and assumed as present. The size of the screen must be known. This information may be supplied over the video-cable or, if this is not possible, it would have to be entered by hand during set-up. Hardware that is able to track the distance between a person and the screen is available, too: It could consist of a simple camera that is calibrated once and then computes the distance from the size of the face. Other options would be triangulating stereo-cameras or ultrasonic sensors. Ambient light sensors are commonly found in TVs or could also be integrated into the distance measuring device. Most smartphones bring all necessary hardware with them as well.

The tests may either be dedicated mini-games (during loading-breaks or inside the actual game in certain situations) or they can be directly integrated into the game. The latter has been proposed previously in US6808267B2, US8931905B2, US9433346B2, US7427135B2, EyeSpy 20/20. A new feature is that the games in our case are no specialised screening games for use in schools or at the ophthalmologist, but that the tests are supposed to be integrated in common games that children actually want to play at home, too. This will increase the frequency of how often the game is played and therefore also how often the test is conducted. The realization of eye tests as a mini-game inside a larger game has to our knowledge never been proposed until now and it offers the unique advantage that the tests can be designed very specific (without the restrictions that are present when implemented into a larger game) and that at the same time the child is being motivated by the larger surrounding game. It is essential that it is a 'real' game that is not identifyable as a vision test in order to provide the necessary motivation for children. As already pointed out in US6808267B2, US9433346B2, further motivation can be provided by rewarding the child with ingame credits for correct actions.

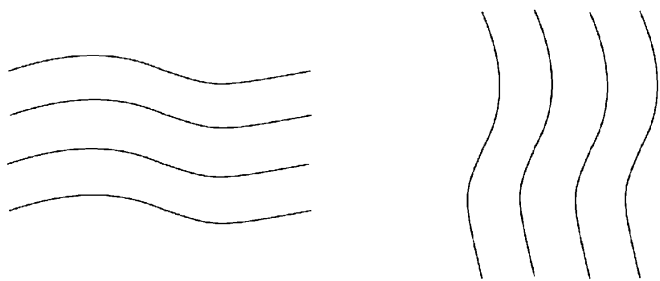

**Figure 2:** A symbol that contains mainly spatial frequencies of a certain direction

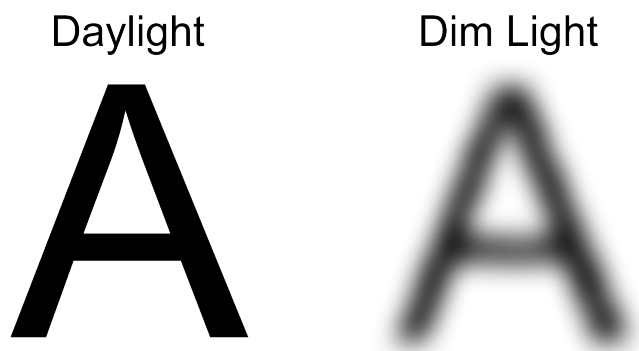

**Figure 3:** The Effect of heavy aberrations with wide open pupil

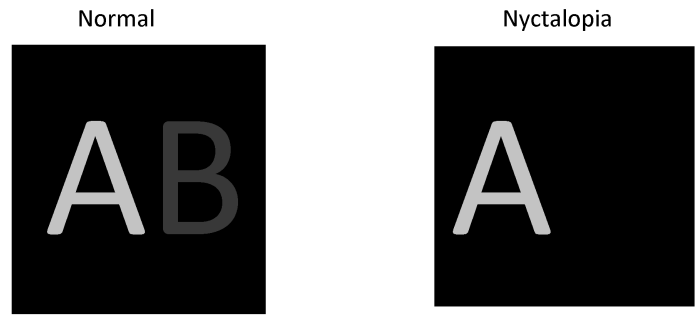

**Figure 4:** With Nyctalopia, the most dim objects can no longer be discerned from complete darkness

In any case the success has to depend on the childrens capability to resolve and recognize certain features on the screen (see examples in Figures 1-5). A mini-game could be game-mechanics-wise completely decoupled from the main game. This would allow for more concrete tests, while the integration into a normal game would require more subtle methods and thus also an evaluation that depends more on statistics than on individual results. For example, in a mini-game, the player could have to find a certain symbol in some static image (possibly a scene from the main game). Another option would be to have moving objects, bearing different labels or having slightly different shape, and the player should only click on the ones with a certain label/shape. For the integrated scenario it would be quite similar: The player would have to interact differently with certain objects depending on their label/shape. They could be integrated into the game for example as power-ups, persons, items, switches or alike.

The impairments that could be monitored are:

- **Far- and shortsightedness:** This is one of the most basic ophthalmic measurements. While some of our ideas are already present e.g. in US6808267B2, US7427135B2 they are in these cases not combined in the way as it is described in the following. Two things can be evaluated:
  1) The average distance the child has to the screen, as it will choose a position from where it can see the screen in a comfortable way. If it is shortsighted it will probably sit closer to the screen while it will sit farther away when it is farsighted.
  2) By investigating what features can apparently be resolved at what viewing distance, one can compute the angular resolution for different viewing distances. This should be similar for all viewing distances. If it is conspicuously low at high or low viewing distances, it could be due to short- or farsightedness again (shortsightedness not really testable on smartphones). The features could simply be letters, or more complex symbols (representing certain mechanisms or items or groups in the game)

- **Color vision capability:** As already stated in US6808267B2, US7427135B2, EyeSpy 20/20, one can estimate what color contrasts the child was able to distinguish. In case of a colorblindness or impaired color vision, this would lead to not being able to detect a feature of a certain color in front of a background of another color, as seen in Figure 1.

- **Astigmatism:** As it is already used in US9433346B2, US7427135B2, one can use features that have high spatial frequencies predominantly in a certain direction as depicted in Figure 2. So when astigmatism is present, in some orientations it will be recognized better than in others.

- **Vision in the dark:** The results that were scored at daylight can be compared to what is scored when the environment is only dimly lit. While e.g. US7427135B2 does utilise a controllable ambient light source in order to check the eye-performance at various ambient light levels, the natural change in brightness in the room where the game is played (at home) has to our knowledge never been used to study the eye in different light-situations before. If fine details can no longer be recognized it may be due to heavy aberrations with wide open apertures, as illustrated in Figure 3. If dim features in dark areas of the image, normally detected by the rods on the retina, can no longer be seen at all, it may be due to nyctalopia. Figure 4 illustrates this condition.

Generally, for each of the above eye conditions one can build symbols that can only be recognized if the condition is not present. If the child is able to see them, everything is ok, if it does not or only seldomly see them, there is probably some impairment. While logging and evaluation of data is widely being used, e.g. in US6808267B2, US9433346B2, EyeSpy 20/20, it usually only incorporates to give the results to the person that is being tested or to transmit the results to a database from where the customer or his doctor can download the data. None of these systems offers the ease of use that a smartphone app included in our idea does. It will allow the parents to monitor their child's status as effortlessly as never before. In case of a detected disease, the parents can straightforwardly send the collected data to an ophthalmologist and arrange a professional examination.

## 3.3 Advantages of the Idea

One advantage is the passiveness, not letting the child know that it is actually doing a vision test. This minimizes cooperation problems, and eases parental monitoring. In addtion, there is the continuity of monitoring that allows for quick identification of eyesight problems.

The main novelties compared to previous systems are:

- Integration into 'real' video games that are regularly played at home

- Integration via mini-games inside the main game

- Comfortable smartphone app which allows the parents to monitor their child's eye health effortlessly

- Test for far-/shortsightedness based on two criteria (angular resolution dependent on viewing distance; average distance)

- Incorporation of ambient light sensor to take into account natural changes in ambient light.

## 3.4 Detectability of the Idea in a Product

In order to allow parents to monitor their children, they have to be made aware of the fact that the game allows for this. This has to be done publically, so it should not be any problem to detect, when other products claim the same functionality. The idea may be included into other products invisibly, too, but without publically claiming so, they will not have additional monetary benefits, making this consideration irrelevant.

# 4 Intelligent Adapting Glasses

Authors: Mathias Unberath, Florian Willomitzer, Florian Schiffers, Maike Stöve

## 4.1 Problem Statement

This idea aims at easing up life for people with progressive sight. Make the need for visual aid glasses with two different optical refraction powers within the same lens for people with progressive sight redundant. Give people with seeing glasses and changing optical refraction power over time the possibility to always have the required adapted seeing lens (adapted diopters). Even adaptation for two different eyes is possible.

Current state of the art technology does not adjust the optical refraction power of the lens upon demand in seeing glasses automatically. This system can also be used in virtual reality glasses as well as augmented reality glasses where users with visual impairments do not have to wear extra glasses. Additionally, the adaptable lens may also provide the ability to magnify close objects like a magnifying glass.

## 4.2 Proposed Method

The idea combines adaptable lenses with the automatic detection of the userŠs line of sight to adjust the optical refraction power of the lens accordingly. The focus depth is determined automatically. This way, the subject carrying the glasses will always have the required seeing aid for any situation.

The glasses incorporate adaptive lenses which are combined with cameras and LEDs in the glasses mount or the lenses itself to track the exact position of the eye's sight. Additionally, there is a distance measurement unit to measure the distance of the eye to the object. Depending on the determined eye position and the distance of line of sight to the object the correct optical refraction power of the lens is set. The user will not require any interaction. The adaptive lenses can be achieved by existing technologies such as liquid crystal lenses. In order to achieve the adjustment electrical power is required. The power supply can either be done by attached batteries or remote power supply techniques to the glass mount.

Possible suggested technical implementations for the self adapting lenses:

- Liquid chrystal lenses

- Variable focus lenses by mechanical pressure

- Piezo chrystals

## 4.3 Advantages of the Idea

The focal length can be adjusted continuously for far and short sighted. The operating point can be changed in case of vision strengths changes of the subject. The adjustment process can be done fully automatically. Optionally there is no user interaction required.

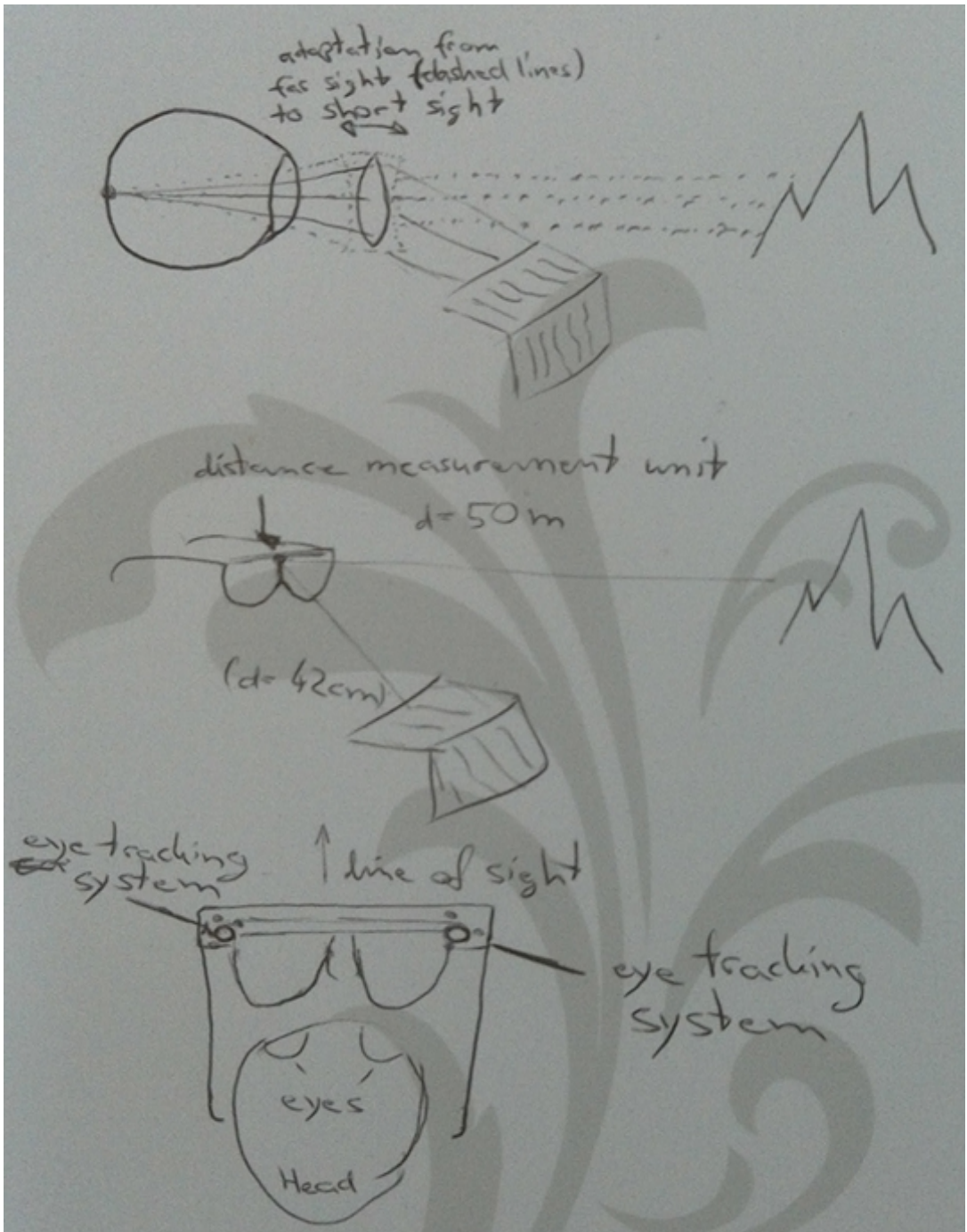

**Figure 5:** Proposed system for intelligent adapting glasses

## 4.4 Detectability of the Idea in a Product

The eye tracking systems in glasses are visible to everyone. Also the distance measure unit is noticeable.

# 5 MD Facebook

Authors: Mathias Unberath, Maike Stöve, Jonas Hajek

## 5.1 Problem Statement

A medical doctor has to be informed about the state of the art methods in his respective field. It is time consuming to keep up with the literature for a practicing doctor. As he already is under time pressure in treating his patients, a detailed search for relevant publications and the following evaluation of the results is often not possible during a regular working day without investing a disproportional amount of time and energy .

Additionally, a doctor might need a second opinion, suggestion or practical value on the diagnosis and treatment of complicated or especially rare diseases.

At the moment, this problems are tackled by m3medical.com, a medical platform providing articles, blogs, news and a doctor forum. As the forum is open for physicians of all specializations, there are not many topics dealing with opthalmology and the community for opthalmology is rather small. Therefore it is hard to search for a special case within the forum and the chance to find valuable information is low. More importantly, using the interface of a conventional internet forum as provided by m3medical is unnecessarily inconvenient. Still it takes a lot of time to type a question or browse for an answer. The same applies to the search for scientific papers and for reading the newsletter. There is also a local restriction of the use of this platform, as a computer and internet conection is mandatory.

Furthermore, it is important that a doctor, who observes an unusual or unexpected progression or manifestation of a disease, is publishing his knowledge. At the moment, publishing such a case is equivalent to writing an article, finding the right journal for publication and being accepted after a time- and energy-consuming review process. This approach is often keeping doctors away from publishing.

## 5.2 Proposed Method

The platform covers two main functionalities. On one hand it has a forum for the interaction between opthalmologists, on the other hand it it has a scientific part containing publications and an easy format for publishing new insights. The forum consists of an easy to understand template to annotate and upload pictures and/or text and/or speech recordings, which will be transcribed automatically by speech recognition software. Coupling the platform and the clinicians network allows to directly upload a picture after a patients visit. A visual and possibly auditive notification will indicate new comments to a forum contribution.

The scientific part will provide the user with a regular newsletters, e.g. daily, weekly or monthly. Every user can specify fields of interest and key words. Based on that, a summary of relevant publications will be presented in the newsletter. Additionally, a easy rating system, e.g. "very exciting", "interesting" or "not relevant" can be used to train the platform for the user and adjust the newsletter keywords. The newsletter can be provided in a written, oral or even audiovisual format. Furthermore. the scientific part contains a search function to browse for publications. An intuitive and easy to use template will enable the doctor to publish newly gained knowledge without undergoing a time consuming review process. To control the quality of forum contributions and publications, a rating system like proposed above should be used.

The platform could be financed by a monthly fee. To motivate opthalmologists for further training, they should be rewarded. Possible types of rewarding are for example:

- Discounts for monthly fee

- CME points

Actions that should be rewarded are for example a highly rated publication or forum contribution. Opthalmologists could also get the opportunity to be rewarded for reading the newsletter for example by optionally doing a short test.

To guarantee, that the platform can be usedvery fast, easy and almost everywhere, it will be compatible to all kinds of mobile devices like smartphones and smartwatches. Additionally, a smartphone can be connected to additional devices and objects like cars, TVs or the radio. Especially mobile devices will have a very simple interface based on speech recognition for posting in the forum. All mobile devices, and devices or objects equipped with speakers will enable the user to listen to the oral format of the newsletter, enabeling him to listen to it in otherwise not actively used gaps of his daily schedule like on the way to work or while doing sports.

The system can be connected to the doctors schedule. This provides the additional possibility to actively ask the user if he wants to listen to the newsletter or wants to submit a new forum contribution. By using machine learning algorithms the system can be trained on the user and adapt to his special needs. For example, the system can learn that a user does not want to listen to the newsletter while driving but likes to listen to it when he is doing sports

## 5.3 Advantages of the Idea

The proposed solution provides major advantages over existing systems. Due to the use of speech recognition, the oral newsletter and the compatibility to many smart devices and objects, the proposed solution can be used very quickly and intuitively. In addition, it can be used almost everywhere. This enables opthalmologists to stay informed about the state of the art in their respective fields during everyday tasks like driving to work or doing sports. The rewarding system will be an additional source of motivation for opthalmologysts to upgrade their education and contribute to the community.

## 5.4 Detectability of the Idea in a Product

The detection of the speech recognition interface and the compatibility to all kinds of mobile devices is obvious.

# 6 Medical Crowd Segmentation

Authors: Florian Schiffers, Maike Stöve, Jonas Hajek

## 6.1 Problem Statement

To obtain features out of different imaging techniques for further processing, regions of interests are required. For example the shape of the eye vessel in a fundus image can be provided by a mask image. Whereas computational algorithms still are inaccurate in distinguishing vessel- and other tissue pixels, humans have proved to perform very well on this task.

On the way to completely automate segmentation tasks, promising possibilities like Deep-Learning approaches require a huge amount of Ground-Truth segmentations. Besides employing people to do these segmentation a different approach is using crowdsourcing. The main goal of crowdsourcing is to split the work and assign it online to volunteers, who will be paid for their contribution. Usually a company (e.g. Amazon with the platform Mechanical Turk) acts as an agent and also charges fees for their services. Considering huge amount of data the main focus is to minimize the cost per segmentation, therefor the described opportunities may be suboptimal. This patent describes an interface for obtaining vessel segmentations, which is cheaper and more independent compared to using permanent employment or external crowdsourcing solutions.

## 6.2 Proposed Method

The proposed interface is a tablet application which focuses on a specific target group of people who enjoy drawing. The idea of the idea is based on adult drawing color books, where adults receive stress reduction due to drawing out printed shapes. Considering the Amazon bestseller list there is a high demand for drawing color books. This trend has also reached the digital field, recognizable by high download numbers for similar digital applications. For example the Android Application ŞAdult Coloring Book PremiumŤ has over 1 million downloads on Google Play Store. It seems reasonable to suppose, that this target group can feel the same stress releasing effect by drawing out image data for segmentation purpose. Because of this positive effect the users may enjoy the task itself and are satisfied with less money for their contribution. Therefor we introduce a user interface for an application which has the following features:

- Connection to database to transfer medical image data

- Painting and erasing function in different brush sizes implemented as a new layer over the image

- Possibility to draw with finger or digitizer

- Guidance for first use to explain the user the segmentation task

- Zoom function to enable accurate work

- On / off button for the painting tool to enable optimal positioning of the image section

- Training / scoring system: The user has to create a certain amount of segmentations before his data will contribute to the output result. For the training data we use ground truth segmentations from physician to compare and evaluate the produced segmentations (e.g. by calculating the mean-square error). The user receives visual feedback of his work by overlaying and highlighting the difference images of the segmentations. Based on his average results a score is assigned to this user in the database. Depending on the amount of available data, many users have to segment the same images and the corresponding output segmentation is calculated by the highest overlap of all segmentations. Users who perform better on the training set will have higher scores and their work is weighted higher for the output segmentation.

- Control system: To ensure high quality after a random number of segmented images the user works again on a Ground-Truth image and his score and feedback is updated.

- Reward system: The user gets money for every accepted segmentation on his account and can get payed out by a suitable payment system. Additional there are prices or bonus payments for users with a very high score.

- Information about the clinical use of the segmentation, to enhance the awareness of the user about the medical issue and the value of his work

The administration of the image and user database can be managed by a web front-end or mobile application. An easy drag and drop upload function of image data enables the upload to physicians. Therefor the user database distinguishes between administrators, image data contributors and users. The idea can be used for segmentation of different medical image data and is not limited to eye images.

## 6.3 Advantages of the Idea

- Cheaper because of omitting third party companies by direct cooperation with the target group and therefor also gaining independence over transferred and stored data.

- High quality results by test section and quality control system during workflow.

- Many contributors by referring to drawing application and stress releasing.

## 6.4 Detectability of the Idea in a Product

The main part of the interface is available to the user, therefore it is easy to check similar approaches from competitors.

# 7 Personalized 3D Model of the Human Eye

Authors: Temitope Paul Onanuga, Mathias Unberath, Florian Doetzer, Jonas Hajek

## 7.1 Problem Statement

The problem solved by the idea is the use of fundus imaging in addition to 3D models of the human eye to create a personalized 3D eye model.

The state of the art is arbitrary 3D models of the human eye. Such models are created using 3D rendering software such as Blender. These models are commonly used for medical doctor training purposes and inclusion in game characters. There are currently no customer available products that make use of personalized eye 3D models.

## 7.2 Proposed Method

Personalized 3D model of the eye consists of a spherical 3D model of the eye, fundus images of the inner eye, and a high resolution image of the outer part of the eye. Recent technology of eye imaging such as D-eye and Peek vision allows taking good quality fundus images with a smartphone camera. Smartphones also usually have good cameras that can produce high quality image of the outer eye. Though this image of the outer eye is intrinsically in 2 dimensions, stereo vision based techniques, e.g taking images of the eye from different positions, or projecting a pattern from the smartphone into the eye, can be used to derive a 3D view of the eye. Similar approach is used by Microsoft MobileFusion application.

These images are then merged together with the 3D eye model to form a personalized 3D representation of the eye. For a more detailed model, fundus images may be combined with OCT images.

Personalized eye models would be offered as a software that allows users to see the whole eye at different perspectives and zoom levels. In addition, these eye models could be 3D printed as a piece of art or included in cartoon figures for children. The kid would be excited to have his favorite character using his own eye.

## 7.3 Advantages of the Idea

- Artistic appeal of the personalized 3D eye model

- Creation of products based on personalized 3D eye model

- This would also subtly encourage people to observe/take eye examinations more frequently, not only when symptoms appear.

## 7.4 Detectability of the Idea in a Product

If a product includes such personalized 3D models then it combines fundus images with a 3D model of the eye.

# 8 Photoacoustic Contact Lens

Authors: Mathias Unberath, Christian Heidorn, Florian Schiffers, Felix Häußler

## 8.1 Problem Statement

The basic idea is to create a contact lens which is able to create/help creating medical images of the eye (perhaps like fundus images or photoacoustic images to measure glucose in blood) to get fast diagnosis during daily grind.

There is a patent of a photoacoustic measurement of the analyte concentration of the eyeball (source: https://www.google.ch/patents/US20080033262?hl=de) but there are no patents of smart lenses which can make diagnosis or images of the eye. The advantage would be that persons do not have to spend time at a physician. Also changes or abnormalities of the retinal region can be better detected, because the smart lens would be able to give information about the inner eye all the time.

## 8.2 Proposed Method

The lens should have its own power supply. This could be realized with small piezo elements or NFC, which are attached onto the lens to generate energy. Another way could be a coil which is loaded through induction from a smart watch.

An Integrated ultrasonic transducer (in example a capacitive micromachined ultrasonic transducer, CMUT) is attached on the lense. The transducer can work as a transmitter. If an AC signal is applied, the vibrating membrane will produce ultrasonic waves to detect blood flow in vessels of the retina. We also can measure the vessel-thickness and see if stenosis exist (see Figure 6).

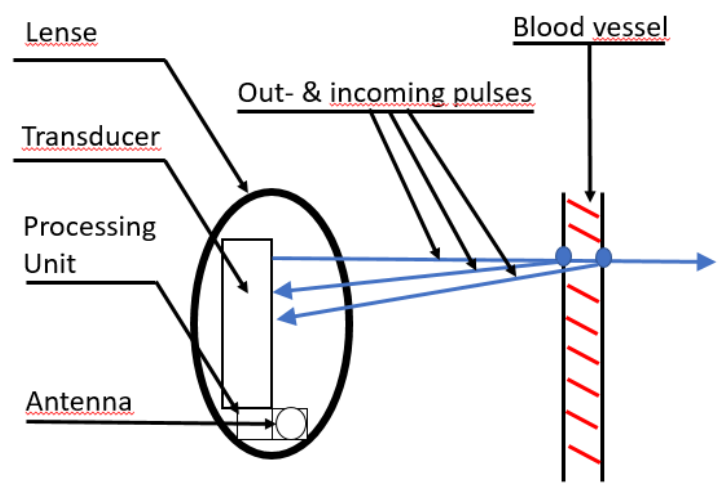

**Figure 6:** Illustration of how the vessel thickness is detected with a transducer. The Transducer turns the electrical energy into ultrasonic pulses. The Transducer measures the time difference between the sent out pulse and the incoming resulting pulses. The results can be sent to a smart watch.

Another option, that is possible, is to create second harmonic generation images. The transducer could also work as a detector (see Figure below). In that case the idea would be to create short laserpulses, so tissue/blood vessels of the retina are heated up. The small ultrasonic waves which are produced could be measured with the transducer.

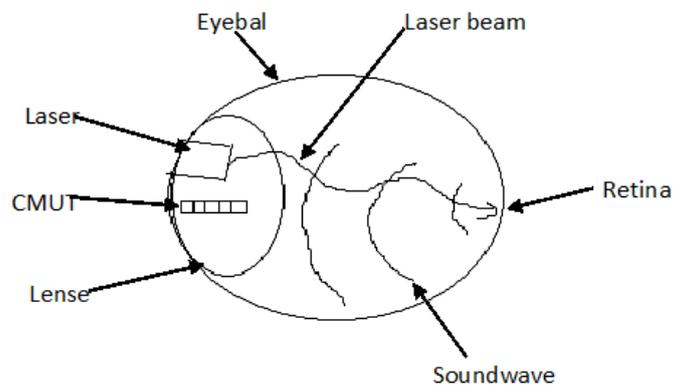

**Figure 7:** Proposed Photoacoustic contact lens

## 8.3  Advantages of the Idea

The eye could be monitored during daily grind without going to the doctor. The ultrasonic measures allow to gain information about the blood cells in the retinal region. Also, glucose measurements, and measurements, if stenosis occurs, are possible. The results can be sent to a smart watch and can be provided for telemedicine, which makes instant diagnosis possible.

## 8.4  Detectability of the Idea in a Product

-

# 9 Power Supply Smart Contact Lens

Authors: Mathias Unberath, Christian Heidorn, Florian Schiffers, Felix Häußler

## 9.1 Problem Statement

This is an idea for power supply a smart contact lenses which could be used in real world or for virtual reality applications. To in example make videos with a cam which is integrated onto the lens one need power supply. Patents for Smart lenses exists (i.e. https://www.google.ch/patents/US20150281411) but there is no method for generating the power mentioned.

## 9.2 Proposed Method

The idea is to generate energy with a piezo element. Through mechanical stress of blinking the eye the energy could be produced and perhaps stored. A piezoelement in form of a strip is attached to the surface of the lens. Through blinking the surface of the eyeball is deformed and therefore the lens is slightly compressed. If the lens is compressed a kind of a small twich on the piezo element will be introduced. Through the piezoelectric effect a small electric current is induced

## 9.3 Advantages of the Idea

The smart lens could act autonomous. No need for a second device. It creates energy by itself.

## 9.4 Detectability of the Idea in a Product

There has to be some experiments done. For example it is not known how a lens is compressed during blinking. It is also not known if the piezo element can create sufficient energy so it acts as a useful power supply.

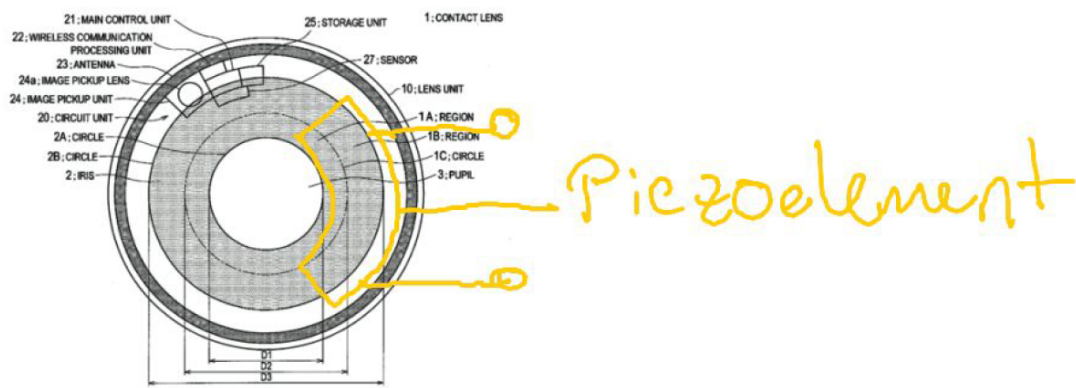

**Figure 8:** Picture from an existing patent of SONY of how a smart contact lense could possible look like. The Piezoelement could be attached on the surface in form of a strip.

# 10 VR-Cornea

Authors: Florian willomitzer, Florian Schiffers

## 10.1 Problem Statement

Measurements of the human cornea can be only performed at the doctor. The alignment of the patient is elaborate and takes a long time. Furthermore, many methods only measure the surface at a few points, not full field.

Prior Art:

- Cornea measurements at the doctor: big tabletop devices

- Measurement methods, that are able to measure reflective surfaces such as the cornea, e.g. Phase Measuring Deflectometry (developed at FAU, Institute of Optics)

- Deflectometric methods, that are implemented on a smartphone.

## 10.2 Proposed Method

Perform deflectometric methods, that are able to measure the human cornea, performed with the smartphone (combination). Mount smartphone in VR glasses. The following components are required:

- Screen, which displays the deflectometric pattern

- Camera (front), which observes the eye

All required components are already implemented in the smartphone. Alternative: Build own device for medical purposes

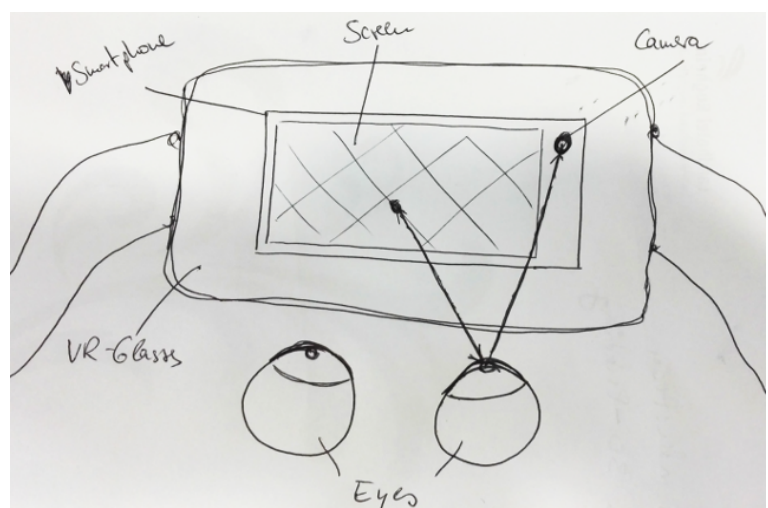

**Figure 9:** Overview of the proposed system

## 10.3  Advantages of the Idea

- No elaborate positioning

- Mobile measurement, no need to go to the doctor

- Full field measurement of the human cornea (not provided by most medical devices).

## 10.4  Detectability of the Idea in a Product

The idea can be identified by the projected pattern.

# 11 Head Mount for Fundus Imaging

Authors: Tobias Geimer, Mathias Unberath, Johannes Wendel, Christian Heidorn, Florian Willomitzer, Gerd Häusler

## 11.1 Problem Statement

Detection of eye deceases is up to the patient to visit a specialized eye doctor on his own accord. Wide-spread screening for eye deceases is currently infeasible as visual impairment is usually not on the check-up list for a general physician. The reason for this is two-fold: First ophthalmic equipment is rather expensive as well as space-consuming for a general practitioner. Second a certain amount of special knowledge is needed to perform ophthalmic imaging with e.g. modalities such as Optical Coherence Tomography (OCT) and Fundus imaging.

In order to enable a visual check-up during a regular visit at the general practice, a compact easy-to-use imaging device is needed, that provides ophthalmic images that can be used for disease classification and early detection without specialized knowledge by the operating general practitioner.

Disadvantages of the state of the art: Current solutions use smartphone LED, therefore, the fundus imaging devices are specific for a particular brand of smartphone light sources. The different lightsource of every smartphone causing potential incompatibilities.

## 11.2 Proposed Method

Setup:

- Head-mount or standing mount with head rest

- Included light source (infrared-light) and corresponding guiding optics

- Adjustable mount for smartphones that allows to adapt the position of the camera to fit every smartphone make

- Connection to synchronize smartphone and mount

The proposed device is a goggle-like encasement similar to existing Virtual Reality (VR) goggles. This could either be used as a low-budget case for smartphone as the imaging device or serve as an active device itself featuring fundus imaging and OCT. Alternatively, a GoPro might be used as the imaging device replacing the smartphone. In conventional systems, the patient is asked to put his chin on a chin rest to fixate the head with respect to the camera. Our approach solves this problem by attaching the imaging system (within the goggle) to the patients head, thus, connecting them in a common rigid system, where the positions of the eye and the system do not change relative to each other.

The device uses an infrared light source for the alignment. It has a movable plate which the smartphone is attached to. The smartphone's sensors should be able to detect infrared light. It could be possible to detect the infrared light with the arbitrary detector of the camera. However, common CCD chips feature an infrared filter blocking of wavelengths above 700nm. For infrared acquisition to be possible the infrared filter needs to be removable. For positioning, it is important that the camera is directly in front of the eye with a small distance. So the smartphone should be fixated inside the system with a band or movable clamps, much akin to a smartphone holder for bicycles. Alignment is achieved via the communication of the smartphone and a motor unit which is used to control the plate. The smartphone takes infrared images of the eye and retina to control the position. An infrared light source will have to be mounted alongside the smartphone to provide illumination. As a result, features like pupil edge and the vessels of the retina can be detected without causing the pupil to close under the influence of the light source. The communication between the smartphone, providing the control images, and the motor unit of the plate ,controlling the position in $x/y$ direction and the angle, can take place via wireless connection. After the ordered position is reached, a signal is sent to the smartphone to take the next picture until an alignment criterion is reached. An acoustic signal is given that the procedure is finished.

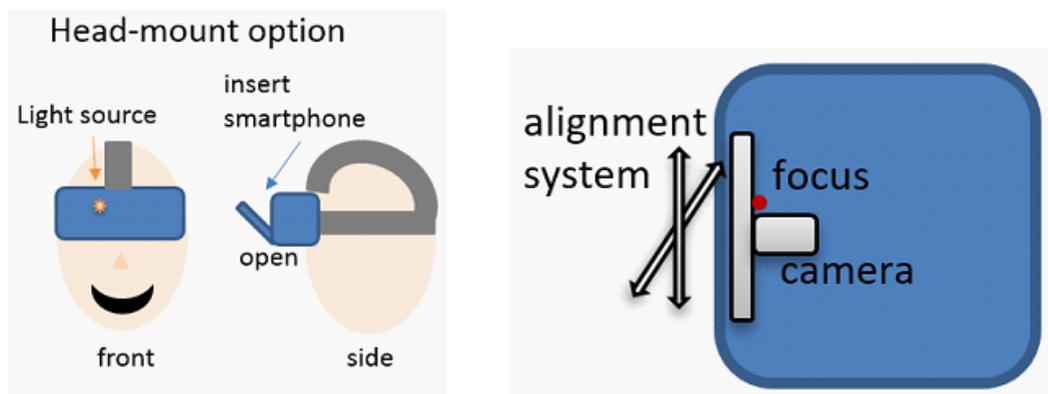

**Figure 10:** Head-mount option and the alignment system to correct angle and x/y direction.

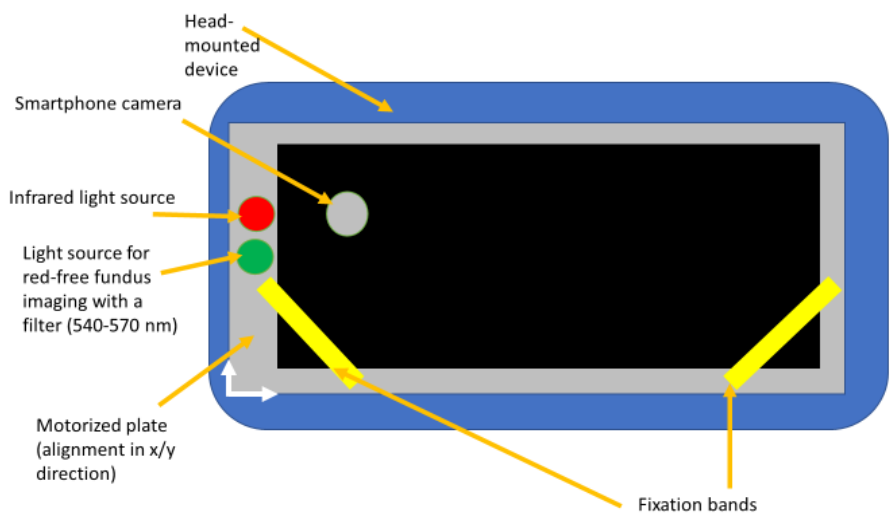

**Figure 11:** System from the patient's point of view. It has two light sources for alignment and the fundus image.

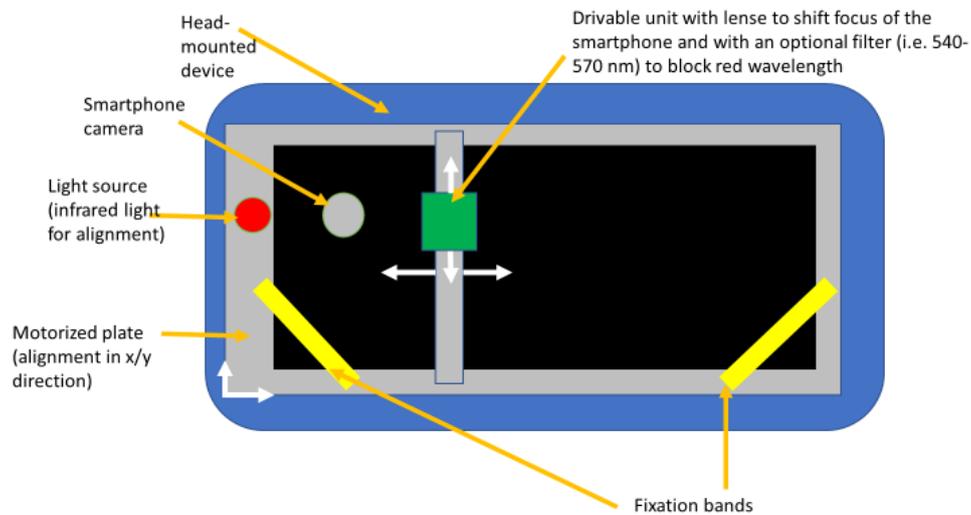

**Figure 12:** System for exploiting the autofocus of the camera with a lens. There is moveable unit to get the lens and the filter for the flash over the camera and its light source (flash) of the smartphone. With that method it is possible to be independent of the type of the smartphone.

On implementation side, processing can be done by a third unit housed in the casing. It takes images from the camera, compares them, sends a movement command to the motor unit, receives a signal, showing that the position is reached, and returns a signal to the camera to acquire a new image. Alternatively, the process can be controlled by a smartphone app omitting the third processing component.

After the alignment process the fundus image is taken. For that there are two possible options for the device setup. The device comes with an integrated light source and beam path, such that the smartphone is only used as a camera. This enables a more robust construction of the illumination pathway and a more flexible mount for the smartphone, that allows the use of virtually any smartphone available on the market. For that the device has two light sources. One with infrared light for the alignment. The other light source includes wavelengths of the visible spectra, with an optional filter (540-570nm) to filter out red wavelengths of light (for better contrast of the blood vessels). If the smartphone is not capable of focusing, for example, to account the patient's refractive error, there is a second option. In front of the Smartphone is an adjustable unit which exploits the cameras autofocus capability. On the principle of direct ophthalmoscopy this unit shifts the focus of the smartphone. The unit has also a filter or a beam splitter to use the camera's flash. However, it needs to filter out specific wavelengths to take the best image possible and to reduce the brightness.

During the acquisition process the patient is presented a focus point to fix his gaze onto. The image is acquired automatically, when alignment is achieved, and illumination is sufficient. Infrared light can be also used for alignment in order to prevent the pupil from closing. The infrared image will serve as the basis for the alignment. Only after the criteria are met, the actual RGB image will be triggered. This is done with the smartphone camera.

The acquisition process involves minimal specialized knowledge by the general practitioner. The device is placed on the patient's head and the process is started via a User Interface on, either an off the shelf desktop PC, or a portable device. After the image is acquired, the doctor can decide to accept it, or to recapture the image. This way, a feedback loop is implemented to check the image quality before the diagnostic image is stored to a data base. Furthermore, this opens up possibilities for telemedicine applications. In this case a potential nurse/layman could position the device, while an expert is actually controls the device and choosing the images from a distance.

Having a screening device of this kind available allows many general practices to acquire many highly comparable images, thus, enabling big data approaches for disease classification. This would increase the early detection rate of common ophthalmic afflictions.

## 11.3  Advantages of the Idea

- Head and imaging system form a common system by attaching the goggle rigidly to the head.

- The goggle is cheaper than a full-blown specialized ophthalmic diagnosis set up.

- Easy to use semiautomatic process requires minimal specialized ophthalmic knowledge.

- Availability at the general practice allows wide-spread screening for eye deceases not limited to eye doctor visits.

- Potential for telemedicine application.

- Same light source for every smartphone allows easier calibration of the camera

- Alignment with infrared light

- Flexible solution enabling use of virtually any smartphone

- Increased diagnostic image quality due to stabilization of head w.r.t. fundus camera

- Robust illumination pathway with potentially increased quality due to fixation

## 11.4  Detectability of the Idea in a Product

The idea provides a new comparatively low-cost method for wide-spread screening of eye diseases during check-up at the general practitioner. It consists of both hardware/housing and software (UI, database, App) components. Thus, any competitor copying the idea is easily identified.